\documentclass[preprint,showpacs,preprintnumbers,superscriptaddress,amsmath,amssymb]{revtex4}
\usepackage{graphicx}
\usepackage{dcolumn}
\usepackage{bm}
\usepackage[american]{babel}
\begin{document}

\title{Probing For New Physics and Detecting non linear vacuum QED effects using gravitational wave interferometer antennas}

\author{G. Zavattini}
\affiliation{INFN, Sezione di Ferrara and Dipartimento di Fisica, Universit\`a di Ferrara, Polo Scientifico, Via Saragat 1 C, 44100 Ferrara, Italy}
\author{E. Calloni}
\affiliation{INFN, Sezione di Napoli and Dipartimento di Scienze Fisiche, Universit\`a ``Federico II", Mostra d'Oltremare, Pad. 19, I-80125, Napoles, Italy}
\noaffiliation

\date{\today}

\begin{abstract}
Low energy non linear QED effects in vacuum have been predicted since 1936 and have been subject of research for many decades. Two main schemes have been proposed for such a 'first' detection: measurements of ellipticity acquired by a linearly polarized beam of light passing through a magnetic field and direct light-light scattering. The study of the propagation of light through an external field can also be used to probe for new physics such as the existence of axion-like particles and millicharged particles. Their existence in nature would cause the index of refraction of vacuum to be different from unity in the presence of an external field and dependent of the polarization direction of the light propagating. The difficulty in reaching a sufficient sensitivity in these experiments to detect QED vacuum non linearities is extreme.
The major achievement of reaching the project sensitivities in gravitational wave interferometers such as LIGO an VIRGO has opened the possibility of using such instruments for the detection of QED corrections in electrodynamics and for probing new physics at very low energies. In this paper we discuss the difference between direct birefringence measurements and index of refraction measurements. We show that in this latter case it is possible to distinguish between various scenarios of new physics in the hypothetical case of detecting unexpected values. We therefore propose an almost parasitic implementation of an external magnetic field along the arms of the VIRGO interferometer and discuss the advantage of this choice in comparison to a previously proposed configuration based on shorter prototype interferometers which we believe is inadequate. 
Considering the design sensitivity in the strain, for the near future VIRGO+ interferometer, of $h<2\cdot10^{-23} \frac{1}{\sqrt{\rm Hz}}$ in the range $40$~Hz $- 400$~Hz  leads to a variable dipole magnet configuration at a frequency above $20$~Hz such that $B^{2}D \ge 13000$~T$^{2}$m/$\sqrt{\rm Hz}$ for a `first' vacuum non linear QED detection.  
\end{abstract}


\maketitle

\section{Introduction}
\subsection{Predicted effects}
Several experimental efforts are underway to detect vacuum magnetic birefringence or direct photon-photon scattering due to non linear QED effects \cite{scatter, HypIn,PRD,Brodin,exawatt,Luiten1,Luiten2,bmv,ni,pugnat}. For photon energies well below the electron mass and for fields much smaller than their critical values, $B \ll B_{\rm crit}={m^{2}c^{2}}/{e \hbar}=4.4\cdot10^{9}$~T, $E \ll E_{\rm crit}={m^{2}c^{3}}/{e \hbar}=1.3\cdot10^{18}$~V/m, these effects are predicted by the Euler-Heisenberg-Weisskopf Lagrangian density correction \cite{QED}
\begin{eqnarray}
L_{\rm EHW}& =& \frac{A_{e}}{\mu_{0}}\bigg[\left(\frac{E^2}{c^2}-B^2\right)^2+7\left(\frac{\vec{E}}{c}\cdot\vec{B}\right)^2\bigg]
\label{LEHW}
\end{eqnarray}
Here the parameter $A_e$ describing the non linearity is given by (S.I. units)
\begin{equation}
A_e=\frac{2}{45\mu_{0}}\frac{\alpha^2 \mathchar'26\mkern-10mu\lambda_e^{3}}{m_{e}c^{2}}=1.32\cdot10^{-24} {\text{~T}}^{-2}
\label{Ae}
\end{equation}
with $\mathchar'26\mkern-10mu\lambda_e$ being the Compton wavelength of the electron, $\alpha={e^2}/{(\hbar c 4\pi\epsilon_0)}$ the fine structure constant, $m_e$ the electron mass, $c$ the speed of light in vacuum and $\mu_0$ the magnetic permeability of vacuum. 

From the Lagrangian density (\ref{LEHW}) combined with the classical electromagnetic Lagrangian density, $L=L_{\rm Class}+L_{\rm EHW}$, one can calculate these two effects. The birefringence induced by a transverse magnetic (or electric) field with respect to the propagation of a laser beam, can be directly derived from the constitutive relations 
\begin{eqnarray}
\vec{D}&=&\frac{1}{\epsilon_{0}}\frac{\partial L}{\partial\vec{E}}\nonumber\\
\vec{H}&=&-{\mu_{0}}\frac{\partial L}{\partial\vec{B}}
\label{constitutive}
\end{eqnarray}
where the fields $\vec{E}$ and $\vec{B}$ in $L=L_{\rm Class}+L_{\rm EHW}$ are the sum of the fields due to the laser and the external field. The result is that the index of refraction depends on the polarization state with respect to the direction of the magnetic field. Indicating with $n_{\parallel}$ and $n_{\perp}$ the index of refraction for light polarized parallel and orthogonally to the magnetic field, respectively, one finds
\begin{eqnarray}
n_{\parallel}=1+7A_{e}B_{\rm ext}^{2}\\
n_{\perp}=1+4A_{e}B_{\rm ext}^{2}
\label{index}
\end{eqnarray}
Vacuum therefore has an index of refraction $n > 1$ in the presence of a magnetic field. Furthermore a birefringence $\Delta n = n_{\parallel}-n_{\perp}$ is also induced \cite{Adler,Iacopini}:
\begin{equation}
\Delta n = 3A_{e}B_{\rm ext}^{2}
\label{birifqed}
\end{equation}

Both photon-photon scattering and the fact that a magnetic field will generate $n>1$, even in vacuum, are connected to the forward scattering amplitude $f(\vartheta=0,E)$ by the relation  (see for example \cite{Newton, Haissinski})
\begin{equation}
n=1+\frac{2\pi}{k^2}Nf(0,E_{\gamma})
\label{n-f}
\end{equation}
where $N$ is the average number density of centers of scattering and $k$ is the photon wave number.

Applying the Lagrangian density (\ref{LEHW}) to photon-photon scattering
of linearly polarized photons, the center of mass forward scattering amplitude of ingoing and outgoing photons all having parallel polarizations, $f^{\rm (QED)}_{\parallel}(0,E_{\gamma})$, and the one in which the two incoming photons have perpendicular polarizations as do the ougoing photons, $f^{\rm (QED)}_{\perp}(0,E_{\gamma})$ are, respectively \cite{Haissinski}
\begin{eqnarray}
f^{\rm (QED)}_{\parallel}(0,E_{\gamma})&=&\frac{32}{45}\frac{\alpha^2\mathchar'26\mkern-10mu\lambda_e}{4\pi}\left(\frac{E_{\gamma}}{m_e c^2}\right)^{3} = \frac{16\mu_{0}}{4\pi\hbar^{2}c^{2}}A_e {E_{\gamma}}^3\\
\label{forwscatt1}
f^{\rm (QED)}_{\perp}(0,E_{\gamma})&=&\frac{56}{45}\frac{\alpha^2\mathchar'26\mkern-10mu\lambda_e}{4\pi}\left(\frac{E_{\gamma}}{m_e c^2}\right)^{3}= \frac{28\mu_{0}}{4\pi\hbar^{2}c^{2}}A_e {E_{\gamma}}^3
\label{forwscatt2}
\end{eqnarray}
where it is apparent that the scattering amplitude is proportional to $A_e$. The authors of \cite{Haissinski} also show that $N$ in equation (\ref{n-f}) is proportional to the energy density of the scatterer field (electric and/or magnetic) and inversely proportional to the photon energy in the center of mass reference frame. From the scattering amplitude one can find the differential cross section
\begin{equation}
\frac{d\sigma_{\gamma\gamma}}{d\Omega}(\vartheta,E_{\gamma})=|f(\vartheta,E_{\gamma})|^{2}
\end{equation}
and the total cross section which depends on $A_{e}^2$. For unpolarized light one finds \cite{DeTollis,Duane,Karplus,Bernard,BernardOld}
\begin{equation}
\sigma_{\gamma\gamma}^{\rm (QED)}(E_{\gamma})=\frac{1}{45^2}\frac{973}{5\pi}\alpha^4\left(\frac{E_{\gamma}}{m_{e}c^{2}}\right)^{6}{\mathchar'26\mkern-10mu\lambda_e^{2}}=\frac{973\mu_{0}^{2}}{20\pi}\frac{E_{\gamma}^{6}}{\hbar^{4}c^{4}}A_{e}^{2}
\end{equation}

In a more general post-Maxwellian description of non linear electrodynamics, the Lagrangian density correction is described by three parameters \cite{Denisov} $\xi$, $\eta_{1}$ and $\eta_{2}$:
\begin{eqnarray}
L_{\rm pM}& =& \frac{\xi}{2\mu_{0}}\bigg[\eta_{1}\left(\frac{E^2}{c^2}-B^2\right)^2+4\eta_{2}\left(\frac{\vec{E}}{c}\cdot\vec{B}\right)^2\bigg]
\label{Lpm}
\end{eqnarray}
In this parameterization $\xi=1/B_{\rm crit}^{2}$, and $\eta_{1}$ and $\eta_{2}$ are dimensionless parameters depending on the chosen model. 
The density (\ref{Lpm}) reduces to (\ref{LEHW}) with $\eta_{2}^{\rm (QED)}=\frac{7}{4}\eta_{1}^{\rm (QED)}=\alpha/(45\pi)$, $\alpha$ being the fine structure constant. 
In this generalization one finds that the birefringence induced by a transverse magnetic field is (to be compared with equation (\ref{birifqed}))
\begin{equation}
\Delta n^{\rm (pM)} = 2\xi(\eta_{2}-\eta_{1})B^{2}
\end{equation}
whereas the forward scattering amplitudes given in expressions (8) and (\ref{forwscatt2}) become
\begin{eqnarray}
f^{\rm (pM)}_{\parallel}(0,E_{\gamma})&=& \frac{8\mu_{0}}{4\pi\hbar^{2}c^{2}}\xi\eta_{1} {E_{\gamma}}^3\\
f^{\rm (pM)}_{\perp}(0,E_{\gamma})&=& \frac{8\mu_{0}}{4\pi\hbar^{2}c^{2}}\xi\eta_{2} {E_{\gamma}}^3
\end{eqnarray}
It is therefore apparent from (\ref{n-f}) how $n_{\parallel}$ depends only on $\eta_{1}$ whereas $n_{\perp}$ depends only on $\eta_{2}$.
Birefringence on the other hand is only sensitive to the difference $\eta_{2}-\eta_{1}$.
Therefore, for example, in the Born-Infeld model \cite{Born} where $\eta_{1}=\eta_{2}$, magnetically induced birefringence is not expected even though $n > 1$.
Therefore ellipsometric techniques alone are not sufficient to determine the two independent quantities $\xi\eta_{1}$ and $\xi\eta_{2}$. On the other hand the direct measurement of $n_{\parallel}$ and $n_{\perp}$ can.

Recently large interferometers designed to search for gravitational waves have reached their predicted sensitivities \cite{virgosens}. This major success has shown that the understanding of such complex systems has reached a very reliable level. Enhanced versions of these systems are expected to reach even better sensitivities \cite{adv} within the next couple of years. The idea of proposing gravitational interferometers for QED measurements has already been suggested in \cite{Denisov} with, though, a few imperfections as discussed in section \ref{apparatus}.
Here we discuss an almost parasitic scheme which could be directly applied to such detectors, not without some difficulty. In particular we will be considering the VIRGO experiment having a better sensitivity at lower frequencies.
\subsection{Probe for new physics}
As discussed by several authors two other important hypothetical effects could also cause $n \ne 1$ in the presence of an external magnetic (or electric) field transverse to the light propagation direction. These can be due either to neutral bosons weakly coupling to two photons called axion-like particles (ALP) \cite{Petronzio,Sikivie,Gasperini}, or millicharged particles (MCP) \cite{Gies,Ringwald}. In this second case both fermions and spin-0 particles can be treated.
\subsubsection{ALP}
Search for axions using laboratory optical techniques was experimentally pioneered by the BFRT collaboration \cite{Cameron} and subsequently continued by the PVLAS effort \cite{PRD}. Initially, this second experiment published the detection of a dichroism induced by the magnetic field \cite{PRL} in vacuum. Such a result, although in contrast with the CAST experiment \cite{CAST}, could have been due to axion-like particles. Subsequently the result was excluded by the same collaboration \cite{NewPVLASeprint,scatter,PRD} after a series of upgrades to their apparatus and almost simultaneously the axion-like interpretation was excluded by two groups \cite{RizzoALP1,RizzoALP2,gammeV} in a regeneration type measurement. Today other such regeneration experiments have confirmed that the original PVLAS signal was spurious. However, the original publication revived interest in the optical effects which could be caused by ALP's and later MCP's.

The Lagrangian density describing the interaction of either pseudoscalar fields $\phi_{\rm a}$ or scalar fields $\phi_{\rm s}$ with two photons can be expressed as (for convenience, written in natural Heavyside-Lorentz units)
\begin{eqnarray}
L_{a} &=& \frac{1}{M_{a}}\phi_{a} \vec{E}\cdot \vec{B}\\
L_{s} &= & \frac{1}{M_{s}}\phi_{s} \left(E^{2}-B^{2} \right)
\label{lagalp}
\end{eqnarray}
where $M_{a}$ and $M_{s}$ are the coupling constants. 

In the pseudoscalar case it is clear from these expressions that in the presence of an external uniform magnetic field $\vec{B}_{\rm ext}$ a photon with electric field $\vec{E}_{\rm \gamma}$ parallel to $\vec{B}_{\rm ext}$ will interact with the pseudoscalar field whereas for electric fields perpendicular to $\vec{B}_{\rm ext}$ no such interaction will exist. For the scalar case the opposite is true: an interaction will exist if $\vec{E}_{\rm \gamma} \perp \vec{B}_{\rm ext}$ and will not if $\vec{E}_{\rm \gamma} \parallel \vec{B}_{\rm ext}$. When an interaction is present, an oscillation between the photon and the pseudoscalar/scalar field will exist.

Therefore for photon energies above the mass $m_{\rm a,s}$ of such particle candidates, a real production can follow. This will cause an oscillation of those photons whose polarization allows an interaction into such particles. On the other hand, even if the photon energy is smaller than the particle mass, virtual production will follow and will therefore cause a phase delay for those photons with an electric field direction allowing an interaction.

The attenuation $\kappa$ and phase delay $\phi$ of those photons with polarization allowing an interaction can be expressed, for both the scalar and pseudoscalar cases, as \cite{Cameron,Petronzio,Sikivie}: 


\begin{eqnarray}
\kappa &=&2\left(\frac{B_{\rm ext}D}{4M_{a,s}}\right)^{2}\left(\frac{\sin x}{x}\right)^{2} \\
\phi &=& \frac{\omega B_{\rm ext}^{2}D}{2M_{a,s}^{2}m_{a,s}^{2}}\left(1-\frac{\sin2x}{2x}\right)
\end{eqnarray}
where, in vacuum, $x=\frac{Dm_{a,s}^{2}}{4\omega}$, $\omega$ is the photon energy and $D$ is the magnetic field length. The above expressions are in natural Heavyside-Lorentz units whereby 1~T $=\sqrt{\frac{\hbar^{3}c^{3}}{e^{4}\mu_{0}}}= 195$~eV$^2$ and 1~m $=\frac{e}{\hbar c}=5.06\cdot10^{6}$~eV$^{-1}$.  The phase delay $\phi$ is related to the index of refraction $n$ by
\begin{equation}
\phi=k\left(n-1\right)D
\end{equation}
Therefore in the pseudoscalar case, where $n^{a}_{\parallel}>1$ and $n^{a}_{\perp}=1$, and in the scalar case, where $n^{s}_{\perp}>1$ and $n^{s}_{\parallel}=1$, one has
\begin{equation}
n_{\parallel}^{a}-1= n_{\perp}^{s}-1= \frac{B_{\rm ext}^{2}}{2M_{a,s}^{2}m_{a,s}^{2}}\left(1-\frac{\sin2x}{2x}\right)
\label{pseudo}
\end{equation}
In the approximation for which $x\ll1$ (small masses) this expression becomes
\begin{equation}
n^{s}_{\perp}-1= n^{a}_{\parallel}-1= \frac{B_{\rm ext}^{2}m_{a,s}^{2}D^{2}}{16M_{a,s}^{2}}
\end{equation}
whereas for $x \gg 1$
\begin{equation}
n^{s}_{\perp}-1= n^{a}_{\parallel}-1= \frac{B_{\rm ext}^{2}}{2M_{a,s}^{2}m_{a,s}^{2}}
\end{equation}
The different behavior of $n^{s}_{\perp}-1$ and $n^{a}_{\parallel}-1$ with respect to $D$ in the two cases where $x\ll1$ and $x\gg1$ is interesting and leaves, in principle, a free experimental handle for distinguishing between various scenarios.

\subsubsection{MCP}
Consider now the vacuum fluctuations of particles with charge $\pm\epsilon e$ and mass $m_{\epsilon}$ as discussed by \cite{Gies,Ringwald}. The photons traversing a uniform magnetic field may interact with such fluctuations resulting in both a pair production if the photon energy $\omega > 2m_{\epsilon}$ and only a phase delay if $\omega < 2m_{\epsilon}$. Furthermore, either fermions or spin-0 charged bosons could exist. Since we are considering the the use of gravitational wave interferometers, only the (real) index of refraction will be considered here.

 - Dirac fermions

Let us first consider the case in which the millicharged particles are Dirac fermions (Df). As derived by \cite{fermion} the indices of refraction of photons with polarization respectively parallel and perpendicular to the external magnetic field have two different mass regimes defined by a dimensionless parameter $\chi$ (S.I. units):
\begin{equation}
\chi\equiv
\frac{3}{2}\frac{\hbar\omega}{m_{\epsilon}c^{2}}\frac{\epsilon e B_{\rm ext}\hbar}{m_{\epsilon}^{2}c^{2}}
\label{chi}
\end{equation}
It can be shown that
\cite{Gies,landaulev}
\begin{equation}
n_{\parallel,\perp}^{Df}=1+I_{\parallel,\perp}^{Df}(\chi) A_{\epsilon} B_{\rm ext}^{2}
\label{ndf}
\end{equation}
with
 \begin{eqnarray}
&I_{\parallel,\perp}^{Df}(\chi)=\\
&=\left\{ \begin{array}{ll}
\left[\left(7\right)_{\parallel},\left(4\right)_{\perp}\right] & \textrm { for  } \chi \ll 1 \\
-\frac{9}{7}\frac{45}{2}\frac{\pi^{1/2}2^{1/3}\left(\Gamma\left(\frac{2}{3}\right)\right)^{2}}{\Gamma\left(\frac{1}{6}\right)}\chi^{-4/3}\left[\left(3\right)_{\parallel},\left(2\right)_{\perp}\right] & \textrm{ for   }\chi\gg 1
\end{array}\right.\nonumber
\label{idf}
 \end{eqnarray}
and
\begin{equation}
A_{\epsilon}=\frac{2}{45\mu_{0}}\frac{\epsilon^{4}\alpha^{2} \mathchar'26\mkern-10mu\lambda_\epsilon^{3}}{m_{\epsilon}c^{2}}
\end{equation}
in analogy to equation (\ref{Ae}).
In the limit of large masses ($\chi\ll1$) this expression reduces to (\ref{index}) with the substitution of $e$ with $\epsilon e$ and $m_{e}$ with $m_{\epsilon}$. The dependence on $B_{\rm ext}$ remains the same as for the well known QED prediction.

For small masses ($\chi\gg1$) the index of refraction now also depends on the parameter $\chi^{-4/3}$ resulting in a net dependence of $n$ with $B_{\rm ext}^{2/3}$ rather than $B_{\rm ext}^{2}$. In both mass regimes, a birefringence is induced:
\begin{eqnarray}
&\Delta n^{Df}&=\left[I_{\parallel}^{Df}(\chi)-I_{\perp}^{Df}(\chi)\right] A_{\epsilon} B_{\rm ext}^{2}=\\
&=&\left\{\begin{array}{ll}
3 A_{\epsilon} B_{\rm ext}^{2}& \textrm{ for  } \chi \ll 1 \\
-\frac{9}{7}\frac{45}{2}\frac{\pi^{1/2}2^{1/3}\left(\Gamma\left(\frac{2}{3}\right)\right)^{2}}{\Gamma\left(\frac{1}{6}\right)}\chi^{-4/3}A_{\epsilon} B_{\rm ext}^{2}& \textrm{ for   }\chi\gg 1
\end{array}\right.\nonumber
\label{deltandf}
\end{eqnarray}

 - Spin-0 charged bosons
 
 Very similar expressions to the Dirac fermion case can also be obtained for the spin-0 (s0) charged particle case \cite{Gies,spin0}. Again there are two mass regimes defined by the same parameter $\chi$ of expression (\ref{chi}). In this case the indices of refraction for the two polarization states with respect to the magnetic field direction are
 \begin{equation}
n_{\parallel,\perp}^{s0}=1+I_{\parallel,\perp}^{s0}(\chi) A_{\epsilon} B_{\rm ext}^{2}
\label{ns0}
 \end{equation}
 with
 \begin{eqnarray}
&I_{\parallel,\perp}^{s0}(\chi)=\\
&=\left\{ \begin{array}{ll}
\left[\left(\frac{1}{4}\right)_{\parallel},\left(\frac{7}{4}\right)_{\perp}\right] & \textrm { for  } \chi \ll 1 \\
-\frac{9}{14}\frac{45}{2}\frac{\pi^{1/2}2^{1/3}\left(\Gamma\left(\frac{2}{3}\right)\right)^{2}}{\Gamma\left(\frac{1}{6}\right)}\chi^{-4/3}\left[\left(\frac{1}{2}\right)_{\parallel},\left(\frac{3}{2}\right)_{\perp}\right] & \textrm{ for   }\chi\gg 1
\end{array}\right.\nonumber
\label{is0}
 \end{eqnarray}
 
As can be seen there is a sign difference in the birefringence $\Delta n$ induced by an external magnetic field in the presence of Dirac fermions with respect to the case in which spin-0 particles exist. This is true for both mass regimes:
\begin{eqnarray}
&\Delta n^{s0}&=\left[I_{\parallel}^{s0}(\chi)-I_{\perp}^{s0}(\chi)\right] A_{\epsilon} B_{\rm ext}^{2}=\\
&=&\left\{\begin{array}{ll}
-\frac{6}{4} A_{\epsilon} B_{\rm ext}^{2}& \textrm{ for  } \chi \ll 1 \\
\frac{9}{14}\frac{45}{2}\frac{\pi^{1/2}2^{1/3}\left(\Gamma\left(\frac{2}{3}\right)\right)^{2}}{\Gamma\left(\frac{1}{6}\right)}\chi^{-4/3}A_{\epsilon} B_{\rm ext}^{2}& \textrm{ for   }\chi\gg 1
\end{array}\right.\nonumber
\label{deltans0}
\end{eqnarray}

\section{Scenario identification}
Assuming a $B_{\rm ext}^{2}$ dependence, at present the best limit on the ratio $\frac{\Delta n}{B_{\rm ext}^{2}}$ induced by a magnetic field is still a factor 5000 greater than the QED value \cite{scatter}:
\begin{eqnarray}
\left(\frac{\Delta n}{B_{\rm ext}^{2}}\right)_{\rm exp.} &<& 1.9\cdot 10^{-20} \textrm{ T}^{-2}\\
\left(\frac{\Delta n}{B_{\rm ext}^{2}}\right)_{\rm QED} &<& 4\cdot 10^{-24} \textrm{ T}^{-2}
\end{eqnarray}
An exhaustive treatment of the possible scenario identification is described in \cite{Gies} considering birefringence and rotation measurements from the various ellipsometric apparatuses. Unfortunately from ellipsometric and rotation measurements alone it is not easy to unambiguously differentiate between the scenario.

On the other hand, the possibility of measuring independently the two values $n_{\parallel}-1$ and $n_{\perp}-1$ would allow the unambiguous identification between the four scenarios if values larger than the QED ones were to be found. In particular the ratio $R=\frac{n_{\parallel}-1}{n_{\perp}-1}$ would suffice together with the determination of whether $n_{\parallel} > n_{\perp}$ or $n_{\parallel} < n_{\perp}$ and whether either $n_{\parallel}=1$ or  $n_{\perp}=1$. Let us examine the different possibilities with the assumption that at least one magnetic field direction has generated a condition where $n > 1$.
\subsection{ALP}
In this hypothesis if $n_{\parallel}>1$ then one must find $n_{\perp}=1$ and the detected effect must be due to a pseudoscalar neutral particle. If, instead, $n_{\perp}>1$ and one finds $n_{\parallel}=1$ then one is observing the effect of a scalar neutral particle coupling to two photons according to (\ref{lagalp}). 

In these conditions the parameters to be determined would be the mass $m_{a,s}$ and coupling constant $M_{a,s}$. The available experimental parameter that one can use for this is the length $D$ of the magnetic field region. Indeed if experimentally one finds $n-1\propto D^{2}$ then $x\ll1$, with $x=\frac{Dm_{a,s}^{2}}{4\omega}$. In this situation the ratio $\frac{m_{a,s}}{M_{a,s}}$ can be determined together with an upper bound on $m_{a,s}$. On the other hand if $x\gg1$ then equation (\ref{pseudo}) tends to a constant independent of $D$ given by $|\Delta n|=\frac{B_{\rm ext}^{2}}{2M_{a,s}^{2}m_{a,s}^{2}}$. In this case the determination of the product $m_{a,s}M_{a,s}$ would be possible together with a lower bound for $m_{a,s}$. 
\subsection{MCP}
In this case there are four different possibilities: Dirac fermion, with $\chi\gg 1$ or $\chi\ll 1$, and spin-0, with $\chi\gg 1$ or $\chi\ll 1$.
Again the ratio $R=\frac{n_{\parallel}-1}{n_{\perp}-1}$ together with the condition $n_{\parallel}>n_{\perp}$ or $n_{\perp}>n_{\parallel}$ can disentangle all four scenarios.

In the Dirac fermion case, it is clear from equations (\ref{ndf}) and (\ref{idf}) that if $n_{\parallel}>n_{\perp}$ then $\chi\ll 1$ whereas if  $n_{\parallel}<n_{\perp}$ then $\chi\gg 1$. Furthermore in these two cases the ratios $R$ would be respectively $R_{\chi\ll 1}=\frac{14}{8}$ and $R_{\chi\gg 1}=\frac{3}{2}$.

In the spin-0 case, from equations (\ref{ns0}) and (\ref{is0}) if $n_{\parallel}<n_{\perp}$ then $\chi\ll 1$ whereas if $n_{\parallel}>n_{\perp}$ then $\chi\gg 1$.
The ratios $R$ for the spin-0 case would be respectively $R_{\chi\ll 1}=\frac{1}{7}$ and $R_{\chi\gg 1}=\frac{1}{3}$.

Finally, for both the fermion and spin-0 cases, the parameters which could be determined for the millicharged particles would be the charge $\epsilon e$ if $\chi \gg\ 1$ and the ratio $\frac{\epsilon e}{m_{\epsilon}}$ if  $\chi \ll\ 1$.

In table \ref{scenario} the various conditions are summarized together with the different quantities which can be determined.
\begin{table}[h!]
\caption{Summary of the ratio $R$ and the parameters which can be determined as a function of $n_{\parallel}$ and $n_{\perp}$.}
\begin{tabular}{c|c|c}
\hline\noalign{\smallskip}
Hypothesis & $n_{\parallel}>n_{\perp}$ & $n_{\parallel}<n_{\perp}$  \\
\noalign{\smallskip}\hline\noalign{\smallskip}
ALP, $x \ll 1$ & $n_{\perp}=1$, $(n_{\parallel}-1)\propto D^{2}$ &  \\
pseudoscalar & det. $\frac{m_{a}}{M_{a}}$, $m_{a}\ll\frac{2\pi\hbar}{\lambda D}$ &  \\
ALP, $x \gg 1$ & $n_{\perp}=1$, $(n_{\parallel}-1)$ indep. $D$ &  \\
pseudoscalar & det. $m_{a}M_{a}$, $m_{a}\gg\frac{2\pi\hbar}{\lambda D}$ &  \\
ALP, $x \ll 1$ &  & $n_{\parallel}=1$, $(n_{\perp}-1)\propto D^{2}$  \\
scalar &  & det. $\frac{m_{s}}{M_{s}}$, $m_{s}\ll\frac{2\pi\hbar}{\lambda D}$  \\
ALP, $x \gg 1$ & & $n_{\parallel}=1$, $(n_{\perp}-1)$ indep. $D$  \\
scalar & & det. $m_{s}M_{s}$, $m_{s}\gg\frac{2\pi\hbar}{\lambda D}$   \\
\hline
MCP, $\chi\gg 1$  &  & $R=\frac{3}{2}$  \\
fermion&  & determine $\epsilon e$  \\
MCP, $\chi\ll 1$ & $R=\frac{7}{4}$ &  \\
fermion & determine $\frac{\epsilon e}{m_{\epsilon}}$ &  \\
MCP, $\chi\gg 1$ & $R=\frac{1}{3}$ &   \\
spin-0 & determine $\epsilon e$ &   \\
MCP, $\chi\ll 1$ &  & $R=\frac{1}{7}$  \\
spin-0&  & determine $\frac{\epsilon e}{m_{\epsilon}}$  \\
\hline
QED & $R=\frac{7}{4}$ & \\
\noalign{\smallskip}\hline
\end{tabular}
\label{scenario}
\end{table}

The power of measuring both $n_{\parallel}$ and $n_{\perp}$ independently is now clear. In the next section we will briefly discuss the magnet configuration which one could imagine to use on a gravitational wave interferometer. We will stress here that the use of a rotating magnetic field would only allow the measurement of $\Delta n = n_{\parallel}-n_{\perp}$ thereby losing in scenario identification power. On the other hand, if a `first' detection of non linear effects in vacuum is the goal, even a rotating magnet could suffice.  
\section{Apparatus and Method}
\label{apparatus}
\begin{figure}[htb]
\begin{center}
{\includegraphics*[width=12cm]{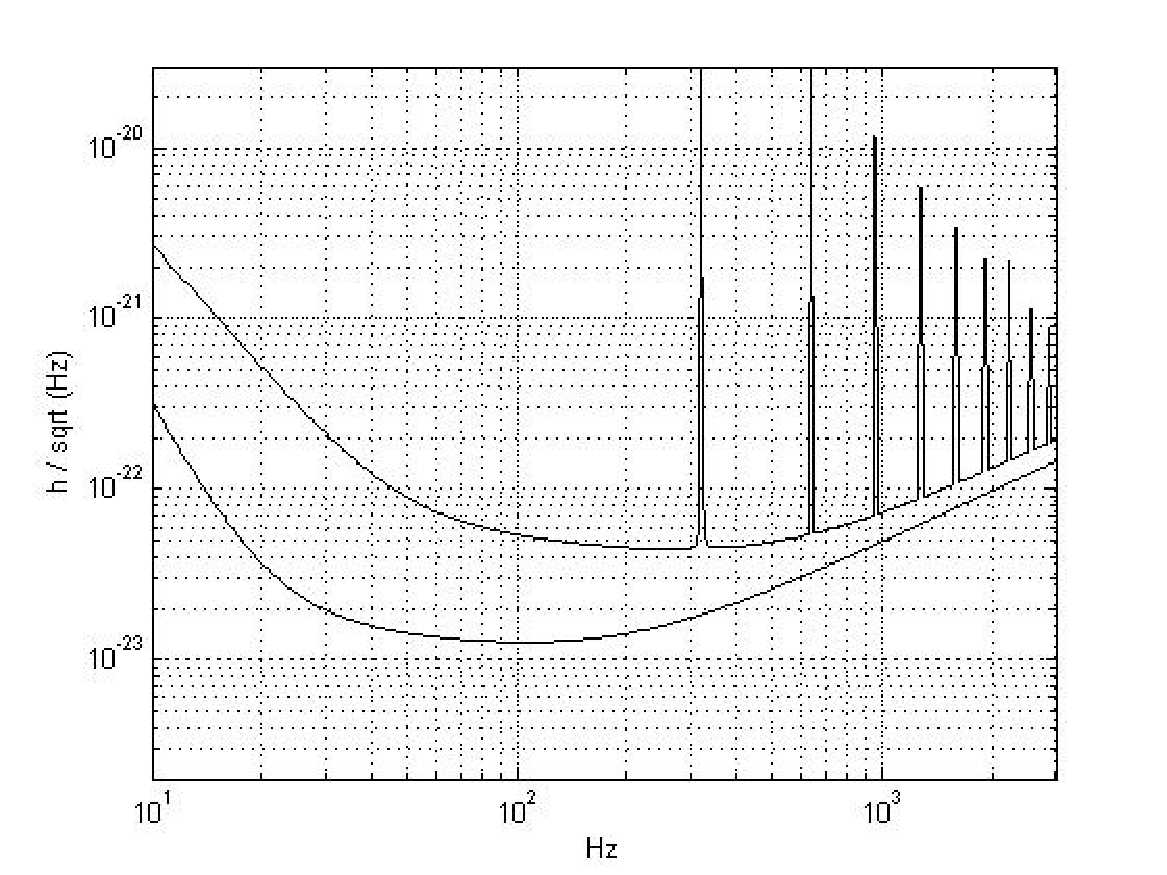}}
\caption{\it Calculated sensitivity curves of the VIRGO (upper) and VIRGO+ (lower) upgraded gravitational interferometers expressed in the strain $h$ as a function of frequency.}
\label{virgosens}
\end{center}
\vspace{-0.6cm}
\end{figure}
Figure \ref{virgosens} shows the 
calculated sensitivities of VIRGO, expressed in the strain $h/\sqrt{\rm Hz}$ \cite{virgosens}, and for the upgraded system called VIRGO+ to be commissioned within the next couple of years. The measured VIRGO sensitivity is about a factor 2 above the calculated VIRGO curve. For VIRGO, between 90 Hz and 1 kHz, the sensitivity in the strain $h=\frac{2\Delta l}{l}$ is better than $10^{-22} \frac{1}{\sqrt{\rm Hz}}$. For VIRGO+ the sensitivity falls below $10^{-22} \frac{1}{\sqrt{\rm Hz}}$ above 25~Hz and is below $2\cdot10^{-23} \frac{1}{\sqrt{\rm Hz}}$ between 40~Hz and 400~Hz. Here $\Delta l$ represents the length variation of one arm, of length $l$, due to the gravitational wave of amplitude $h$. Therefore $\Delta L = 2\frac{4F}{\pi} \Delta l=\frac{4F}{\pi}hl$, where $F$ is the finesse of the Fabry-Perot cavity constituting each arm, represents the difference of {\em effective} path length of the beams recombining at the output of the interferometer. 

In a recent paper  \cite{Denisov}, discussing the use of interferometry for similar measurements, it has been argued that the use of long-arm interferometers is not sensible and shorter interferometers would be preferable. The main argument is that the sensitivity to the difference in effective path lengths of two interferometer arms $\Delta L$ is essentially the same for short or long interferometers and that, as a consequence, it is more efficient to use a short interferometer with suitably high number of bounces N with respect to a long interferometer with a small number of bounces. In the same paper the number of bounces is generally considered instead of the finesse of the Fabry-Perot cavity constituting each arm. There is no difference, in principle, at this level but, for sake of precision, we will use the finesse $F$ in the following discussion.

We must point out that unfortunately the previous arguments are misleading and lead to wrong conclusions. Let us consider the length variation, $\Delta l$, of one arm of the interferometer, such that $\Delta L = \frac{8F}{\pi}\Delta l$. 
It is true that prototype interferometers with short arms have demonstrated a sensitivity in $\Delta L$ comparable to long interferometers, but only with a finesse of the same order of magnitude; it is not demonstrated at all that the sensitivity to the physical arm-length variation $\Delta l=\frac{\pi}{8F}\Delta L$ will scale as $F$. To be explicit: if the sensitivity $\Delta L$ of an interferometer is limited by the physical motion of the mirrors,  the sensitivity on the physical signal $\Delta l_{\rm s}$ will not benefit at all by increasing the finesse of the cavities because both the signal $\Delta l_{\rm s}$ and the noise $\Delta l_{\rm n}$ will be enhanced by the finesse in the same way. This is the case, for example, of the motion due to thermal noise of the suspension system and/or the mirrors themselves. Indeed in long-arm interferometers, at low frequencies (below 100 Hz) the sensitivity is limited by the thermal noise of the suspensions.

We also note that since for gravitational wave detection $\Delta L = \frac{8F}{\pi} \Delta l = \frac{8F}{\pi} l h$ (note that $\Delta l$ is the length variation of the single arm) then by using the same considerations as in \cite{Denisov} it would also seem that one could use short-arm high finesse interferometers. But the above considerations have lead to the construction of long-arm low finesse systems.

In particular the proposed experiment \cite{Denisov} make use of finesses of the order of $10^{5}$. It is questionable, and not at all demonstrated experimentally, that the sensitivity will scale with $F$ with an increment of $F$ of about three orders of magnitude. On the contrary we discuss the feasibility of an experiment with the current (and very next future) long arm interferometers, whose sensitivity is well established. Furthermore, as it will be shown later, the use of long arms allows to increase the actual signal, $\Delta l_{\rm s}$, because the physical length of the interacting region can be made in principle much longer, of the order of hundreds meters; it also allows to maintain well separated (Km apart) the region of production of magnetic field, that can be placed in the middle of the arm, from the region of detection of signal, at the arm ends, where the mirrors and the read-out electronic are located, thus minimizing spurious coupling. As a conclusion, in our opinion, the use of short arm interferometers could be of interest, provided that deep studies on noise will be performed to prove feasibility but, contrary to the statement of \cite{Denisov}, present long arm interferometer are at least as well suited to investigate post maxwell electrodynamics.

To compare the post-maxwell effect and gravitational wave interferometers sensitivity let us consider the difference of length variation of the two arms  $\Delta l_{\rm s} = \Delta l_1 - \Delta l_2$. Upon considering the effect of the gravitational strain on one arm $\Delta l = \frac{1}{2} h l $ and
that the gravitational wave acts in a differential way on the two arms, the effect of a differential arm variation $\Delta l_{\rm s}$ is equivalent to the strain $h_{\rm eq} = \frac{\Delta l_{\rm s}}{l}$.
If along one arm of the interferometer there is a region with index of refraction $n>1$ of length $D$ it will generate a variation in the one pass optical path length  $\Delta l_{\rm s}$ equivalent to the gravitational strain: 
\begin{equation}
h_{\rm eq}=\frac{\Delta l_{\rm s}}{l}=\frac{(n-1)D}{l}
\end{equation}
Given a sensitivity $h_{\rm sens}$, and an index of refraction $n\ne 1$ the necessary integration time $T$ for a signal to noise ratio of unity is
\begin{equation}
T=\left(\frac{h_{\rm sens}}{h_{n}}\right)^{2}=\left(\frac{h_{\rm sens}l}{(n-1)D}\right)^{2}
\end{equation}
Let us first consider the predicted QED vacuum fluctuation contribution to $n$. Given a time dependent magnetic field with direction parallel to the light polarizzation, $B(t)_{\rm ext} = B_{\rm 0}\cos(2\pi\nu t)$, then 
\begin{equation}
n_{\parallel}-1 = 7A_{e}B_{\rm ext}^{2} = \frac{7}{2}A_{e}B_{\rm 0}^2\left[1+\cos(4\pi\nu t)\right]
\end{equation}
It must be noted that the index of refraction, which depends on $B_{\rm ext}^{2}$, will therefore vary at twice the frequency of the magnetic field.
Therefore if $\nu$ is the frequency variation of the magnetic field and $h(2\nu)_{\rm sens}$ is the sensitivity at $2\nu$ then
\begin{equation}
T=\left(\frac{h(2\nu)_{\rm sens}l}{(n_{\parallel}-1)D}\right)^{2}=\left(\frac{2}{7}\frac{h(2\nu)_{\rm sens}l}{A_{e}B_{\rm 0}^{2}D}\right)^{2}
\end{equation}
Considering a reasonable integration time $T=10^{6}$~s and a time dependent index of refraction $n$ at a frequency such that the sensitivity $h_{\rm sens}<2\cdot10^{-23} \frac{1}{\sqrt{\rm Hz}}$, as expected for VIRGO+, this would require that
\begin{equation}
B_{\rm 0}^{2}D\ge h(2\nu)_{\rm sens}\frac{2}{7}\frac{l}{A_{e}\sqrt{T}} = 13\textrm{ T$^2$m}
\end{equation}
Similarly, the measurement of the index of refraction $n\ne 1$ for a magnetic field perpendicular to the polarization would result in
\begin{equation}
B_{\rm 0}^{2}D\ge h(2\nu)_{\rm sens}\frac{2}{4}\frac{l}{A_{e}\sqrt{T}} = 23\textrm{ T$^2$m}
\end{equation}
Finally, for a magnet in which the field strength $B_{\rm ext}=B_{\rm 0}$ is constant but the field direction is rotated around the beam direction at a frequency $\nu$, only $\Delta n = n_{\parallel}-n_{\perp}$ could be detected (again at $2\nu$) and the necessary magnet would need to satisfy
\begin{equation}
B_{\rm 0}^{2}D\ge h(2\nu)_{\rm sens}\frac{2}{3}\frac{l}{A_{e}\sqrt{T}} = 30\textrm{ T$^2$m}
\label{b2d}
\end{equation}

This is an impressive magnet system especially if it needs to be modulated at several tens of hertz where the sensitivities of gravitational antennas are best. A modular system of magnets could be installed so as to begin by improving existing limits of $n-1$ hence on the existence of ALP candidates and/or MCPs with a shorter field length.

As a comparison to ellipsometric apparatuses the expression for the necessary $B_{\rm 0}^{2}D$ for detecting magnetic vacuum birefringence with ellipsometric experiments is \cite{scatter}
\begin{equation}
B_{\rm 0}^{2}D\ge \frac{\psi_{\rm sens}\lambda}{2F}\frac{1}{3A_{e}\sqrt{T}}
\end{equation}
where $\psi_{\rm sens}$ is the ellipticity sensitivity, $\lambda$ is the wavelength of the light and $F$ is the finesse of the optical cavity.
Comparing this to the gravitational interferometer values in (\ref{b2d}) a sensitivity comparison of the two techniques can be made with regards to birefringence measurements.

\begin{equation}
 \frac{\psi_{\rm sens}\lambda}{4F} \iff h_{\rm sens}l = \Delta l_{\rm sens}
\end{equation}
where $\Delta l_{\rm sens}$ is the absolute length variation spectral density of a single interferometer arm.

Today $F$ can be as high as $4\cdot10^{5}$ with $\psi_{\rm sens} = 10^{-7}$~1/$\sqrt{\rm Hz}$ @ 10 Hz \cite{private} resulting in $\frac{\psi_{\rm sens}\lambda}{4F}=1.4\cdot10^{-19}$ m/$\sqrt{\rm Hz}$ @ 10 Hz to be compared to $\Delta l_{\rm sens}=h_{\rm sens}l \le 6\cdot10^{-20}$ m/$\sqrt{\rm Hz}$ between 40~Hz and 400~Hz for VIRGO+.

\subsection{Magnet constraints}
A detailed description of a magnet is not in the aim of this paper but some constraints on the magnet system will be presented. In the PVLAS experiment the dipole magnet is a superconducting magnet 1~m long with a 5.5 T field resulting in $B_{\rm 0}^{2}D=30$~T$^2$m. Therefore such magnets exist concerning the field strength and length. The difficulty in implementing a magnet on a gravitational wave interferometer is the bore hole it would need. The VIRGO interferometer has a beam waist $w_{0}=2.5$~cm at each of the entrance mirrors of the cavities. Furthermore the vacuum beam pipe is equipped with baffles about 35~cm in diameter at regular intervals of about 20~m to block stray light which would otherwise spoil the sensitivity. Smaller diameters would result in beam clipping hence noise generation in the interferometer itself. Finally the beam pipe itself is 1.2~m in diameter and is made of metal.

Three possible magnet implementations could be considered ordered according to how invasive it would be:
\begin{enumerate}
\item Magnet surrounding the whole beam pipe for a length $D$.
\item Magnet surrounding a portion of beam pipe where a narrower section is introduced. Such a section could have a diameter of 0.5 meters and continue to be equipped with the existing baffles. 
\item Magnet could be directly inserted inside the vacuum pipe. This option would allow the smallest bore hole, although it would need to be at least 35~cm in diameter and, obviously, vacuum compatible to pressure levels of $10^{-8}$ mbar.
\end{enumerate}
Furthermore in the first two cases a non metal tube would be ideal to avoid Foucault currents.

In the above discussion it is clear that to measure $R$ both a field parallel and perpendicular to the polarization of the light are necessary. To obtain such fields again several different configurations are possible:
\begin{enumerate}
\item A single magnet capable of generating fields both vertically and horizontally.
\item Separate magnet systems along the same interferometer arm generating each a field in a single direction.
\item Separate magnet systems for each field direction. The magnets generating the different field directions would be placed on each of the interferometer's arm.
\item A single magnet generating a field in one direction but mechanically rotatable around an axis parallel to the beam direction.
\end{enumerate}
Depending on how the current in the magnets is driven, all of the above solutions would allow the measurement of $n_{\parallel}$, $n_{\perp}$ and $\Delta n$ except for solution 4 which could not measure directly $\Delta n$.

In any of the above cases the energy involved and the technical difficulties are not negligible especially if the field needs to be modulated at several tens of hertz necessary to match the optimal sensitivity of existing interferometers. 

\section{Conclusion}
It is known that non linear QED effects in vacuum and probing for new physics with an external transverse magnetic field would induce indices of refraction $n\ne 1$. We have discussed that an important difference lies between measuring birefringences, $\Delta n$, and measuring independently the indices of refraction parallel, $n_{\parallel}$, and perpendicular, $n_{\perp}$, to the magnetic field. This difference lies in the capability to successfully distinguish between the various scenarios.
 Indeed, the ratio $R=\frac{n_{\parallel}-1}{n_{\perp}-1}$ together with the sign of $\Delta n$ and lastly whether either $n_{\parallel}=1$ or $n_{\perp}=1$, allows the unambiguous determination of the scenario in the case of a signal larger than the predicted vacuum QED magnetic birefringence. We have also discussed how the existing full-scale interferometric gravitational wave antennae offer a unique opportunity to perform such fundamental tests with an almost parasitic integration of a transverse magnetic along one or both of the two interferometer arms. 

\end{document}